\documentclass[twocolumn, trackchanges]{aastex63}
\shorttitle{Discovery of Probable X-Ray Emitting CSM in the Synchrotron-Dominated SNR RX~J1713.7-3946}
\shortauthors{Tateishi et al.}
\graphicspath{{./}{figures/}}

\begin{document}

\title{Possible Detection of X-Ray Emitting Circumstellar Material in the Synchrotron-Dominated Supernova Remnant RX~J~1713.7-3946}


\

\author{Dai Tateishi}
\affiliation{Graduate School of Science and Engineering, Saitama University, 255 Shimo-Okubo,Sakura, Saitama, 338-8570, Japan}

\author{Satoru Katsuda}
\affiliation{Graduate School of Science and Engineering, Saitama University, 255 Shimo-Okubo,Sakura, Saitama, 338-8570, Japan}

\author{Yukikatsu Terada}
\affiliation{Graduate School of Science and Engineering, Saitama University, 255 Shimo-Okubo,Sakura, Saitama, 338-8570, Japan}
\affiliation{Japan Aerospace Exploration Agency, Institute of Space and Astronautical Science, Sagamihara, Kanagawa, Japan}

\author{Fabio Acero}
\affiliation{Laboratoire AIM, IRFU/SAp – CEA/DRF – CNRS – Université Paris Diderot, Bât. 709, CEA-Saclay, Gif-sur-Yvette Cedex,
France}

\author{Takashi Yoshida}
\affiliation{Yukawa Institute for Theoretical Physics, Kyoto University, Kyoto 606-8502, Japan}

\author{Shin-ichiro Fujimoto}
\affiliation{National Institute of Technology, Kumamoto College, Kumamoto 861-1102, Japan}

\author{Hidetoshi Sano}
\affiliation{National Astronomical Observatory of Japan, Mitaka, Tokyo 181-8588, Japan}


\

\begin{abstract}

We report on a discovery of an X-ray emitting circumstellar material knot 
inside the synchrotron dominant supernova remnant (SNR) RX~J1713.7-3946.  
This knot was previously thought to be a Wolf-Rayet star (WR~85), but we realized that it is in fact $\sim$40$^{\prime\prime}$ away from WR~85, indicating no relation to WR~85.  We performed high-resolution X-ray spectroscopy with the Reflection Grating Spectrometer (RGS) on board {\it XMM-Newton}.  
The RGS spectrum clearly resolves a number of emission lines, such as N Ly$\alpha$, O Ly$\alpha$, Fe XVIII, Ne X, Mg XI, and Si XIII.  
The spectrum can be well represented by an absorbed thermal emission model with a temperature of $k_{\rm B}T_{\rm e} = 0.65\pm 0.02$\,keV.  
The elemental abundances are obtained to be ${\rm N/H} = 3.5\pm 0.8{\rm \left(N/H\right)_{\odot}}$, ${\rm O/H} = 0.5\pm0.1{\rm \left(O/H\right)_{\odot}}$, ${\rm Ne/H} = 0.9\pm0.1{\rm \left(Ne/H\right)_{\odot}}$, ${\rm Mg/H} = 
1.0\pm0.1{\rm \left(Mg/H\right)_{\odot}}$, ${\rm Si/H} = 1.0\pm0.2{\rm \left(Si/H\right)_{\odot}}$, and ${\rm Fe/H} = 1.3\pm0.1{\rm \left(Fe/H\right)_{\odot}}$.  The enhanced N abundance with others being about the solar values allows us to infer that this knot is circumstellar material ejected when the progenitor star evolved into a red supergiant.  The abundance ratio of N to O is obtained to be $\rm N/O = 6.8_{-2.1}^{+2.5}\left(N/O\right)_{\odot}$.  By comparing this to those in outer layers of red supergiant stars expected from stellar evolution simulations, we estimate the initial mass of the progenitor star to be $15\, \rm M_{\odot} \lesssim \rm M \lesssim 20\, \rm M_{\odot}$.

\end{abstract}



\section{Introduction} \label{sec:intro}

RX~J1713.7-3946 is a shell-type supernova remnant (SNR) discovered by the 
{\it ROSAT} All-Sky Survey\, \citep{WR85}.  Its angular diameter is about 1$^\circ$, and the distance to the remnant is estimated to be $\sim$1\,kpc 
based on radio and optical observations\, \citep[e.g.,][]{Fukui_2003, fukui_2012, Leike_2020}.  
X-ray measurements of the expansion velocity suggest that this SNR is $1{,}580 \mathchar`- 2{,}800$ years old\, \citep{Tsuji_2016, Acero_2017}.  This age, combined with its location, led to a possible relation to SN~393 recorded in Chinese history books\, \citep{SN393}.

This SNR can be seen in wide-band wavelengths from radio\,\citep{Slane_1999, Ellison_2001, Lazendic_2004, Sano_2020} to very-high-energy gamma rays \,\citep{Muraishi_2000, Enomoto_2002, Aharonian_2004, Aharonian_2006b, Aharonian_2007, Abdo_2011, HESS_2018}.
In the X-ray, synchrotron radiation from relativistic electrons is very strong, while thermal radiation is extremely weak\,\citep{Koyama_1997, Slane_1999, Uchiyama_2003, Cassam_2004, Tanaka_2008, Acero_2009, Higurashi_2020, Tanaka_2020}.
Therefore, this SNR is thought to be one of the most important objects to study particle acceleration in SNRs.

A neutron star, 1WGA~J1713.4-3939, is reported at the location close to the geometric center of the SNR.  
Its surface temperature, radius, and luminosity obtained by X-ray spectral analysis are similar to those of the Central Compact Objects (CCO), a subclass of young neutron stars.  
Also, the hydrogen column density matches that for the SNR.  
These facts strongly suggest that it is the CCO of SNR~RX~J1713.7-3946, and that this SNR results from a core-collapsed SN\,\citep{Cassam_2004}. 

The progenitor star of the SNR has been inferred by several methods.  
By assuming that the SNR shell has the same size as the cavity created by the stellar wind, \,\cite{Cassam_2004} estimated the initial mass of the progenitor star to be $12$-$16\, {\rm M_{\odot}}$.  
Recently, thermal X-ray emission originating from SN ejecta was found at the center of the SNR\,\citep{Katsuda_2015}.  
The X-ray measured chemical abundances, combined with those expected from SN nucleosynthesis models, allowed for estimating an initial mass of the progenitor star to be $\lesssim 20\ {\rm M_\odot}$.  
With this initial mass, Type~\mbox{I\hspace{-.1em}IP}~SN is feasible.  
However, the velocity of explosive ejecta expected with Type \mbox{I\hspace{-.1em}IP} SN $\left(3{,}000 \mathchar`-4 {,} 000\ {\rm km\,s^{-1}} \right) $ is significantly less than the observed mean velocity of $\sim 5 {,} 900\ {\rm km\,s^{-1}}$.  
Rather, the mean velocity is closer to the expansion speed for Type Ib/c SN.  
To resolve this contradictory result, \citet{Katsuda_2015} proposed that the progenitor star is a close binary with an initial mass of $\lesssim 20\ {\rm M_ \odot}$, in which binary interactions removed a massive H envelope.  

The purpose of this study is to constrain the initial mass of the progenitor star of SNR~RX~J1713.7-3946.  It is known that chemical abundances and velocity of the circumstellar material (CSM) depends on the initial mass of a progenitor star \,\citep[e.g.,][]{Chiba_2020}.  Therefore, it is possible to estimate the initial mass by precisely measuring these parameter from precise spectroscopic analysis of thermal X-rays emitted from the CSM.  
The paper proceeds with section~2 describing the observations and data reduction, section~3 analysis and results, section~4 providing some discussion, and ends with the summary.

\begin{figure}
    \centering
    \includegraphics[width=8cm,clip]{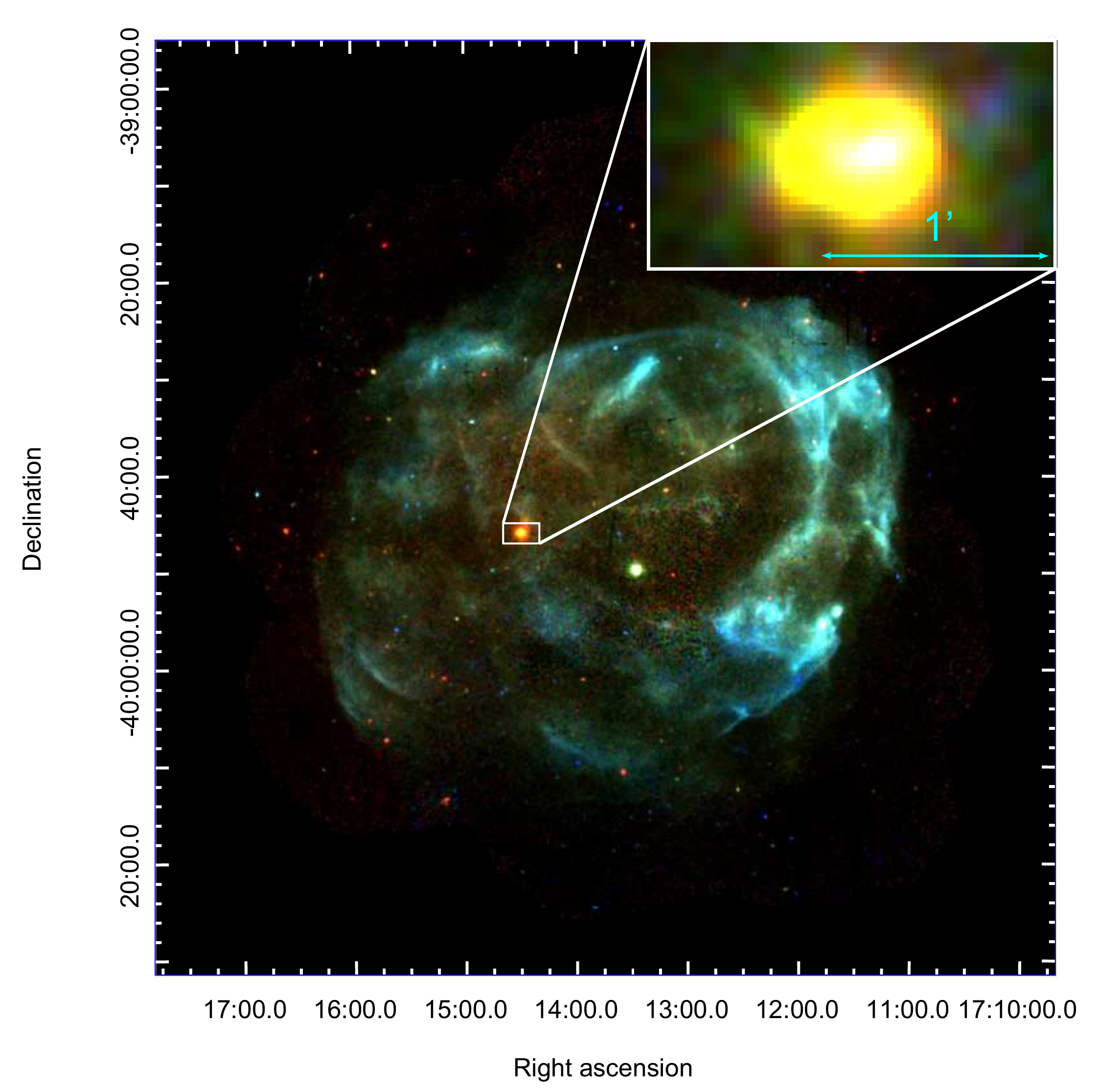}
    \caption{X-ray false color image of RX J1713.7-3946 observed by {\it XMM-Newton} MOS and pn. Red, green and blue represent 0.2-1.0 keV, 1.0-2.0 keV, and 2.0-4.5 keV energy band, respectively.}
    \label{fig:all}
\end{figure}

\section{Observation and Data Reduction} \label{sec:obs}

From 2017 to 2018, we (PI: F.\ Acero) performed a large observation campaign of RX~J1713.7-3946 with {\it XMM-Newton}, aiming at a deep X-ray observation of the entire SNR.  
The angular size of RX~J1713.7-3946 is so large, a radius of $\sim$30$^\prime$, that 10 pointings are required to cover the entire remnant.  
All the raw data have been reduced with the {\it XMM-Newton} SAS v16.1.0. 

Using all the {\it XMM-Newton}/MOS data, we created a false colored image (0.2--4.5 keV) as shown in Figure \ref{fig:all}.  
This figure shows that most of the part is dominated by high energy X-rays (2.0--4.5 keV), whereas a knotty feature in the east of the SNR center is enhanced in low-energy X-rays, and is readily distinguished from the other regions.  
This structure was previously associated as a Wolf-Rayet star, WR~85, when it was found by {\it ROSAT} observations\,\citep{WR85}.
However, our observations with {\it XMM-Newton}/MOS, which has a higher angular resolution, revealed that the knot is offset from WR~85 by $\sim33''$, which is much larger than the pointing accuracy of {\it XMM-Newton} (less than $1''$ in root mean square:\,\citealt{XMM_pointing}).  
In addition, it turned out to be a diffuse source, which will be proved in the following analysis.  
Therefore, as tested in detail in the following sections, we considered that the knot is not WR~85 itself.

In this paper, we focus on this peculiar knot, hereafter K1.  Fortunately, it is located near the on-axis position of one pointing taken on 2017 August 30 (Obs.ID 0804300801), providing us with data from the Reflection Grating Spectrometer (RGS).  Thus, we use the RGS as a primary instrument.  In general, the RGS is not useful for diffuse 
sources like SNRs, because it does not have a slit.  However, if the angular size of the target is small enough (less than a few arcmin) and is brighter than its surroundings, it is possible to obtain high-resolution spectra from the RGS.  The diameter of our target is small enough ($0.8'$) to obtain a high-resolution spectrum with the RGS.  We also analyze data obtained with the MOS to support the RGS analysis and obtain additional information.  
 Fortunately, the data are almost free from background flares due to 
solar soft protons.  We did not have to exclude bad time periods for the RGS, whereas we excluded short time periods for the European Photon Imaging Camera (EPIC) Metal Oxide Semi-conductor (MOS).  The resultant effective exposure times are 46.7\,ks/46.7\,ks 
and 44.6\,ks/44.4\,ks for the RGS1/2 and MOS1/2, respectively. 

\section{Analysis and Results} \label{sec:analysis}

To reveal the nature of K1, we investigate its detailed morphology and measure its chemical abundances.

\subsection{Morphological Analysis of K1}
\begin{figure}
    \centering
      \includegraphics[width=8cm]{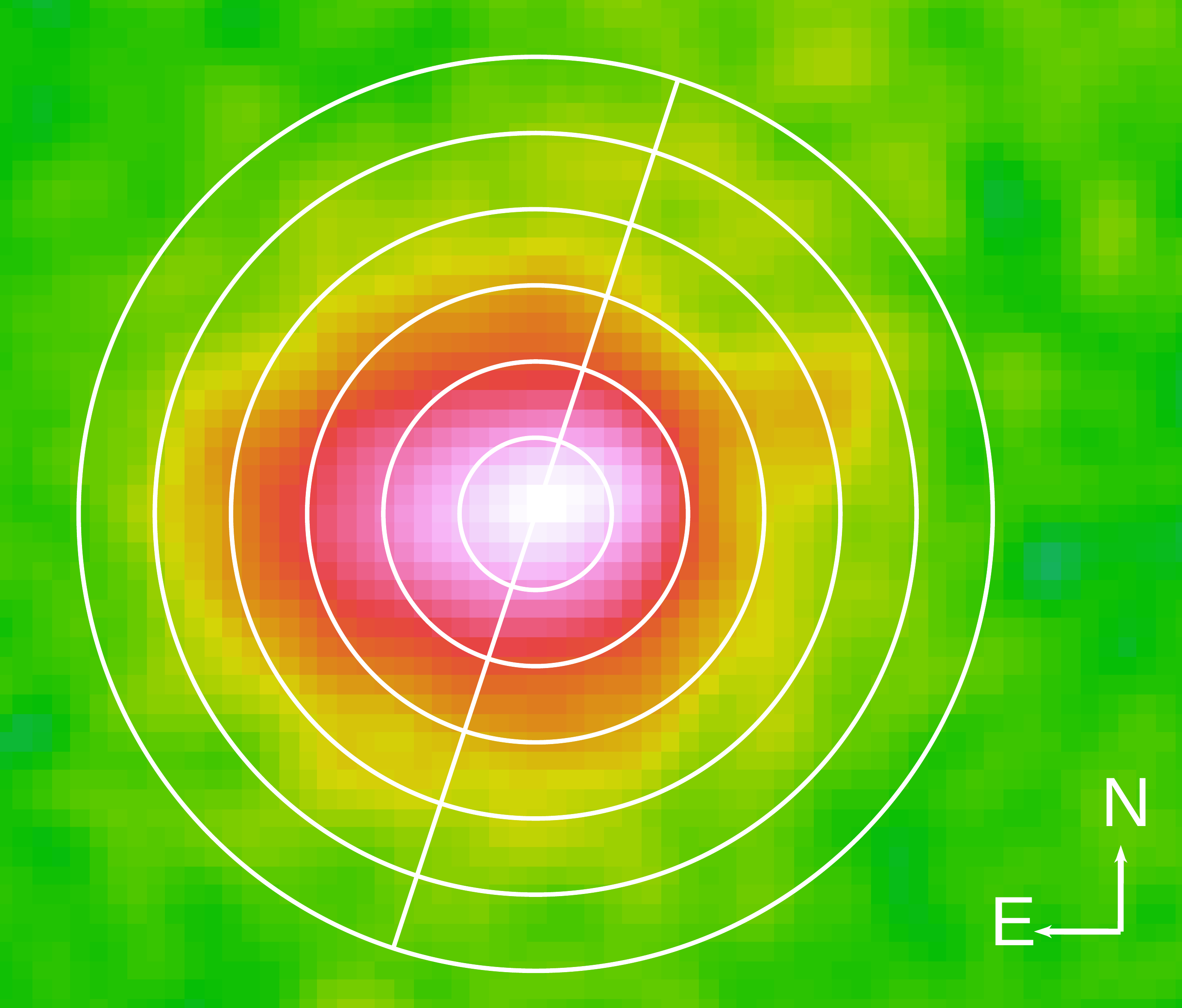}
     \caption{
     Analysis area of azimuthal distribution of K1.
     The extraction region overlaid on the {\it XMM-Newton}/MOS image in the energy band of 0.45-5.0\, keV.
     }
     \label{fig:radial_reg}
\end{figure}

As can be seen in Figure \ref{fig:radial_reg}, K1 seems to be a diffuse source, and have an elliptic shape with head (west) and tail (east) features.  
We quantitatively investigate the morphology as follows.
\begin{enumerate}
    \item
    Make an image of K1 in the energy band of 0.45-5.0 keV using {\it XMM-Newton}/MOS.  
    Determine the center of K1 where the brightest pixel exists.
    \item
    Divide K1 in the east and west with respect to the center of K1.  The division line is taken to be perpendicular to the line connecting the central compact object (1WGA J1713.4-3939) and the center of K1.
    \item
    Calculate surface brightnesses (photons per pixel) in concentric annuli with increasing radii of $8''$ in the east and west sides, respectively, and obtain brightnesses as a function of angular distance from the center of K1.
    \item
    Compare the brightness profile in the east and west parts with the Point Spread Function (PSF) of the MOS.
    The PSF was generated at the energy of 1 keV and at the position of K1 using the {\it XMM-Newton} SAS {\it psfgen} tool.
\end{enumerate}
\begin{figure}
      \includegraphics[width=8cm]{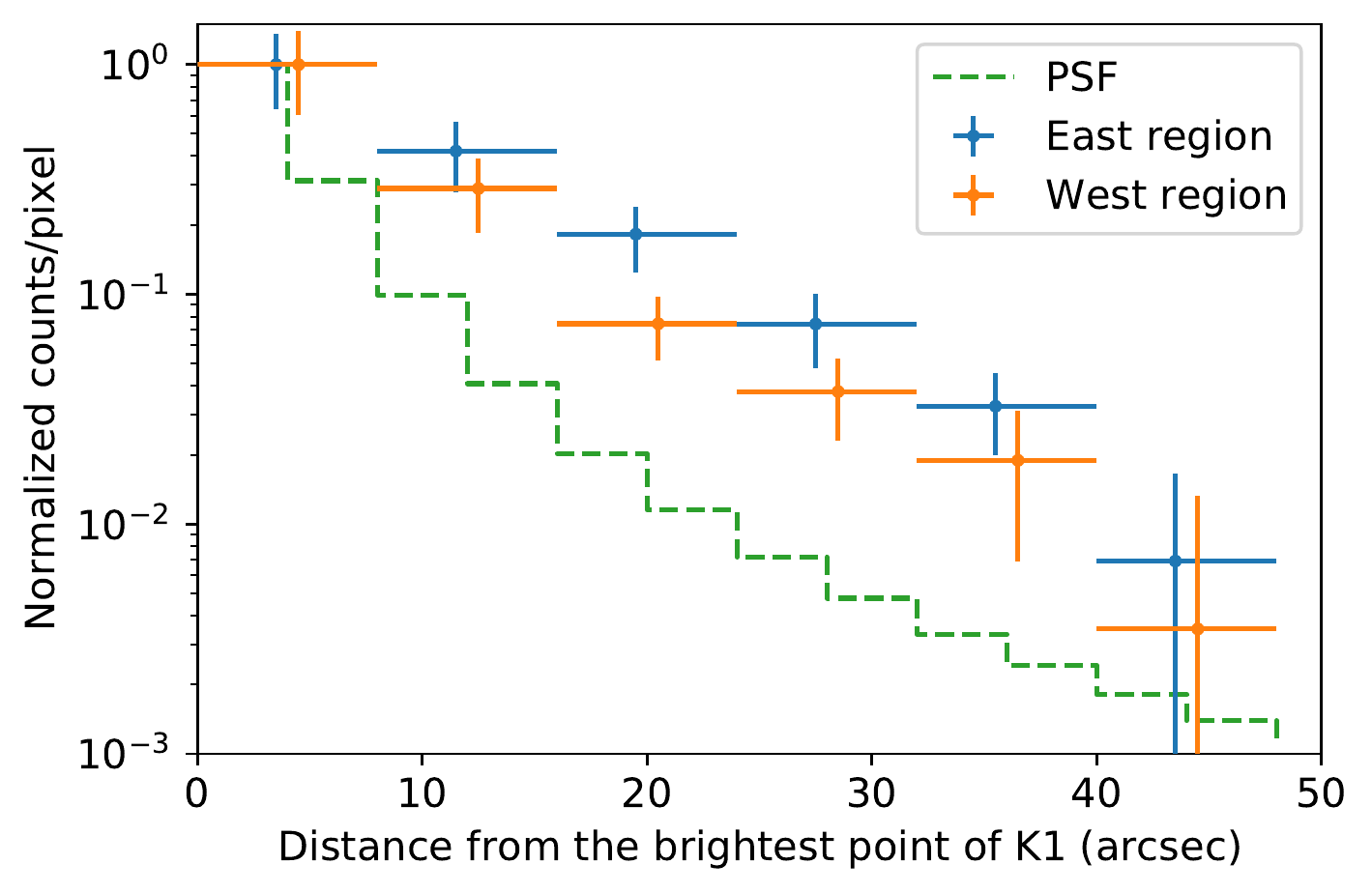}
      \caption{
        The number of photons per unit pixel in each azimuthal direction of K1.
        The number of counts in each region are shown in Table \ref{tab:radial_counts}.
        The green dashed line represent the PSF at the position of K1, while blue and orange line represent observation result of east and west region respectively.
        The values was normalized to the number of photons per pixel at the innermost region.
        The error bars represent 1$\sigma$ statistical error.}
      \label{fig:radial}
\end{figure}
The results are shown in Figure \ref{fig:radial}.
The number of counts in each region are shown in Table \ref{tab:radial_counts}.
It is clear that X-ray emission in both east and west extend larger than the PSF.  
In addition, the east part shows a slower decline outward than the west, supporting the head- and tail-like structures along the west-east direction.

\begin{figure}
    \centering
    \includegraphics[width=8cm,clip]{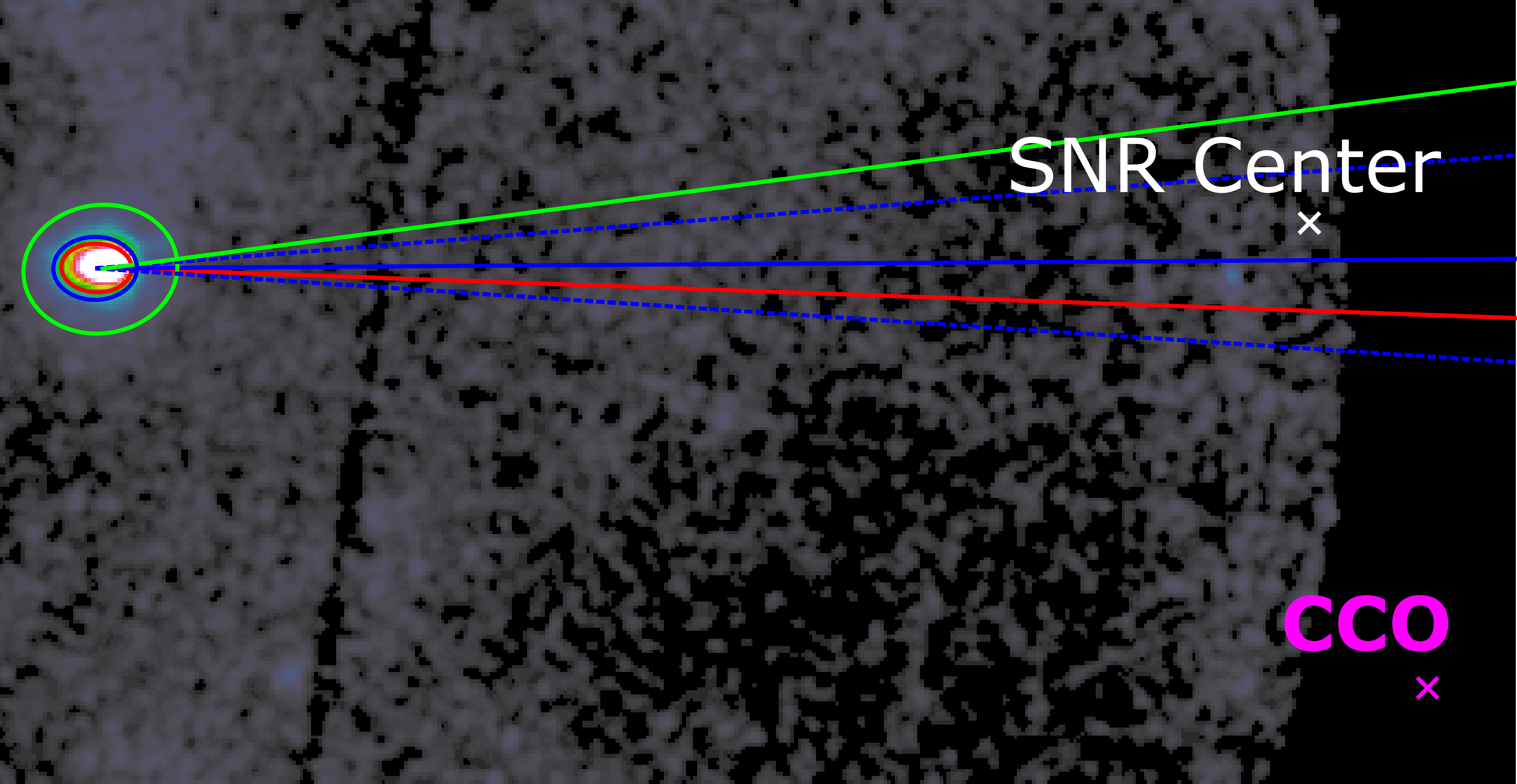}
    \caption{The ellipse fitting result of K1. 
    The green, blue and red lines represent the results from signal to noise ratio of 2, 5, and 8, respectively.
    The solid line in each color represent the best-fit. The blue dashed lines represent the direction in $1\sigma$ statistical error.
    The white point represent the nominal center of the remnant\, \citep{Tsuji_2016}, and magenta point represent the coordinate of CCO object\,\citep{Slane_1999}.}
    \label{fig:ellipse}
\end{figure}

To quantitatively measure the direction of K1's elongation, we fitted the edge of K1 with an ellipse.
The edges were defined by intensity contours of K1 at various signal-to-ratios ranging from 2--10.
The best-fit major axes are shown as solid lines in Figure \ref{fig:ellipse}.
The green, blue, and red lines represent results from contour levels with signal-to-noise ratios of 2, 5, and 8, respectively.
The blue dashed lines represent the $1\sigma$ statistical uncertainty on the case with the signal-to-noise ratio of 5.
It is clear that the major axis points to the SNR center.  

\begin{deluxetable*}{ccc}
\tablenum{1}
\label{tab:radial_counts}
\tablecaption{
Number of counts used in morphological analysis
}
\tablewidth{0pt}
\tablehead{\colhead{Distance from the brightest point of K1} & \colhead{Counts in east region} & \colhead{Counts in west region}}
\startdata
\hline
$0''-8''$&825 & 843\\
$8''-16''$&1219 & 844\\
$16''-24''$&960 & 464\\
$24''-32''$&700 & 445\\
$32''-40''$&547 & 414\\
$40''-48''$&405 & 349\\
\hline
\enddata
\end{deluxetable*}

\subsection{Spectral Analysis}
\label{sec:ana_spec}
We analyze X-ray spectra obtained by the RGS and the MOS simultaneously.  

\begin{figure*}
    \centering
    \includegraphics[width=18cm,clip]{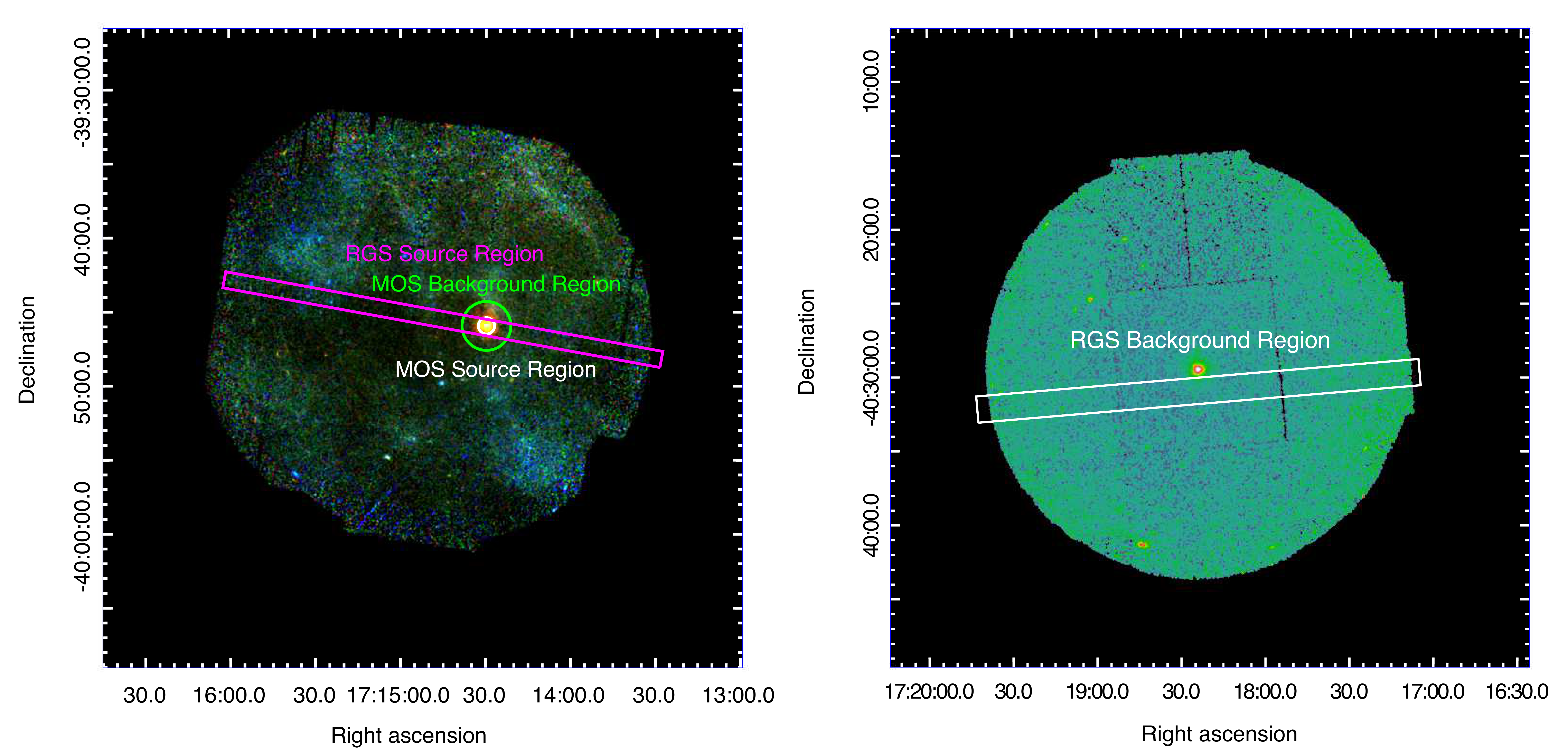}
    \caption{
    Spectral extracted regions for source and background data. 
    The figure on the left shows RGS source region (magenta), MOS source region (white) and MOS background region (green).
    The figure on the right represent RGS background region.
    }
    \label{fig:rgs_region}
\end{figure*}

\begin{figure}
    \centering
    \includegraphics[width=8cm,clip]{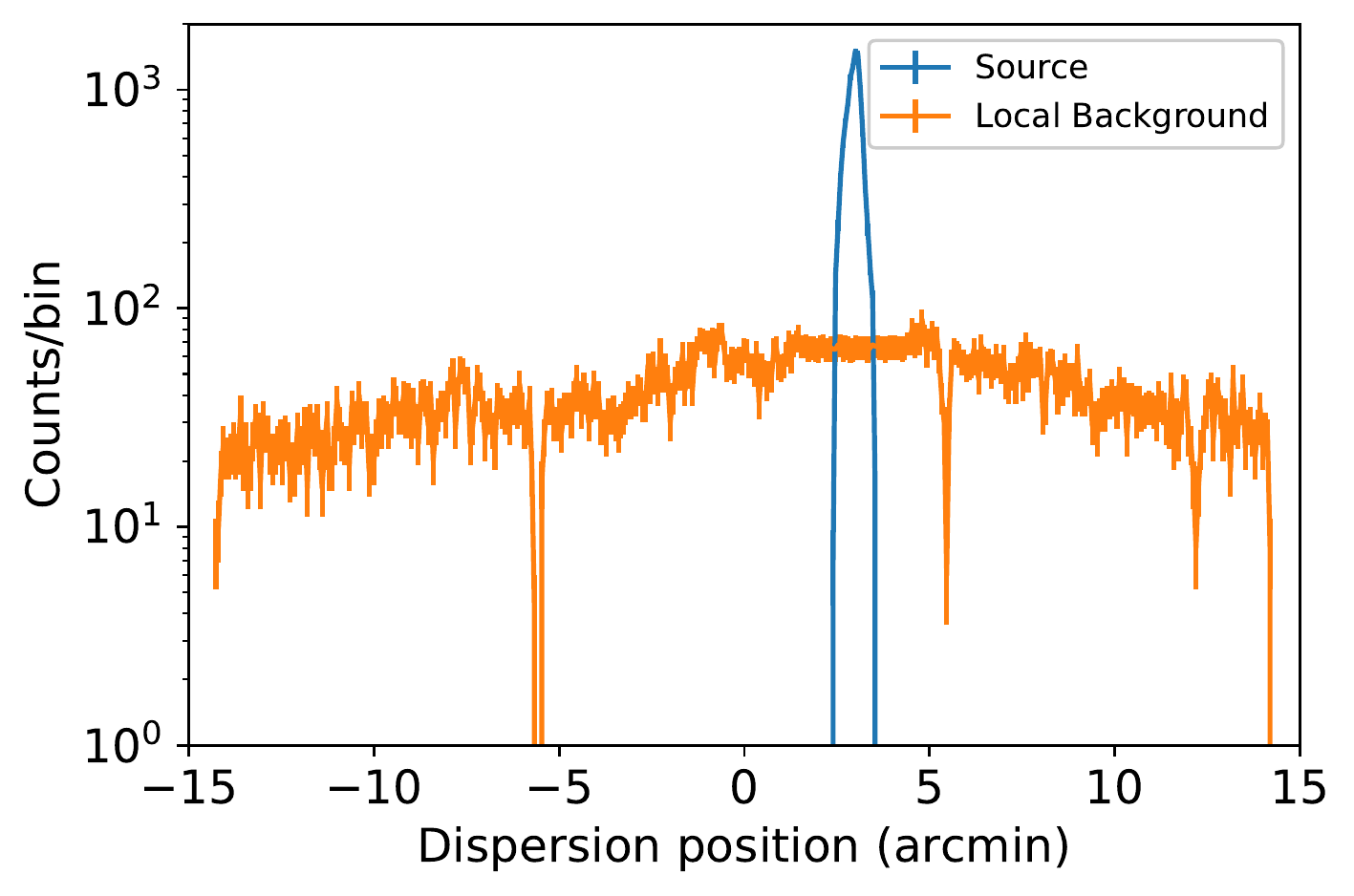}
    \caption{
    Emission profile which we used to smooth the RMF.
    Orange line shows profile which we used to smooth RMF for local background radiation, while blue line represent profile which we used to smooth RMF for knot emission.
    }
    \label{fig:smoothed_rmf}
\end{figure}

Source and background extraction regions are shown in Figure \ref{fig:rgs_region}.
For RGS analyses, we extracted the 1st-order RGS spectra from the area of 
$1.1'$ wide centered on K1.
Note that we use a different observation data (Obs.ID: 0605160101, Exposure: 66.2\,ks/66.4\,ks for MOS1/2 respectively.), aiming 
at a low-mass X-ray binary, 1RXS~J171824.2-40293, to subtract the non X-ray background (NXB).  
Since K1 is diffuse, we convolve the response 
matrix file (RMF) produced by the SAS tool {\it rgsproc} with {\it XMM-Newton}/MOS images using the {\it rgsrmfsmooth} tool.  
In this procedure, we input two MOS images for K1 and its surrounding emission, both in energy range of 0.45--2.0\,keV.
The former and latter MOS images are modified to focus on K1 and to exclude K1, respectively.  
The resultant emission profiles used to smooth the RMF are shown in Figure \ref{fig:smoothed_rmf}.

For our MOS analysis, we extracted spectra from a circular region with a radius of $33''$.
The background region was extracted from the surrounding ring-shaped region with a radius of $99''$, as shown in Figure \ref{fig:rgs_region}.
The total number of counts used in spectral analysis are shown in Table \ref{tab:counts}.

\begin{figure*}
    \centering
    \includegraphics[width=18cm,clip]{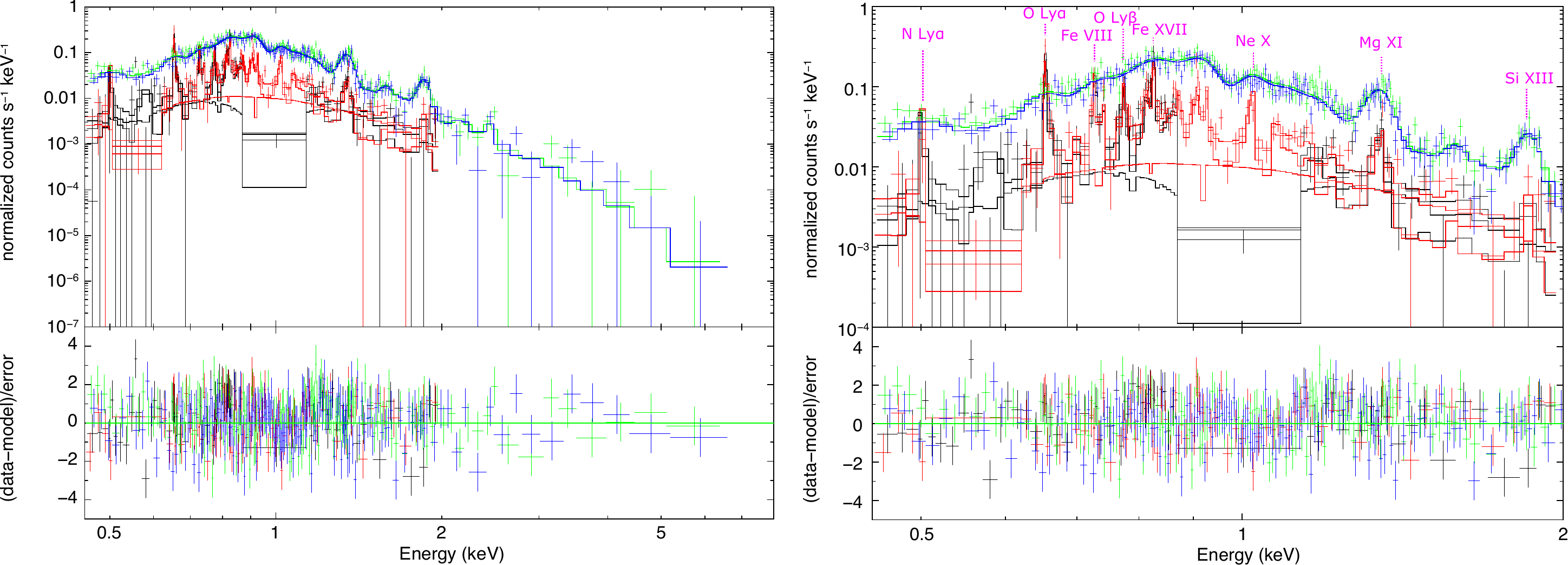}
    \caption{
    The spectrum of K1 observed by {\it XMM-Newton}.
    Black, red, green, and blue lines represent X-ray observed by RGS1, RGS2, MOS1, and MOS2 respectively. 
    The lines represent the best-fit spectra obtained from VPSHOCK model.
    The lower panel shows residual of data to best-fit model.
    The right panel represent the same plot as left but zoom version on the low energies.
    The spectral discontinuities for RGS1 and 2 are due to in-flight loss of CCD.
    Magenta line represent line identification.
    }
    \label{fig:best_fit}
\end{figure*}

We present the RGS and MOS background-subtracted spectra in Figure \ref{fig:best_fit}.  
The RGS spectrum clearly shows a number of emission lines, such as N Ly$\alpha$, O Ly$\alpha$, Fe XVIII, O Ly$\beta$, Fe XVII, Ne X, Mg XI, Si XIII, for the first time from K1.  
Since several emission lines were observed, it is clear that the X-rays emitted from K1 are thermal emission.
We fitted the MOS and RGS spectra simultaneously with a thermal emission model using XSPEC version 12.9.1\,\citep{xspec}.  
Specifically, we adopt an absorbed (TBabs:\,\citealt{tbabs}), single temperature plane-parallel shocked plasma model (VPSHOCK).
We also took into account possible broadening of the emission line by multiplying a gaussian smoothing function (gsmooth).
The overall expression of thermal emission model is TBabs$\times$ gsmooth$\times$ VPSHOCK.
This model implicitly assumes that line widths are linked to that of the brightest line, in this case O~Ly$\alpha$.
VPSHOCK model is parameterized by plasma temperature, elemental abundance, lower and upper limit of ionization timescale, redshift, and emission measure.
For the solar abundance ratio, we used \citet{Lodders_2009}.
It is well known that RX~J1713.7-3946 radiates intense synchrotron X-rays.  
To explain such a ``local X-ray background", we add an absorbed (TBabs) power-law component, which is represented with photon index and normalization.

In addition, we fitted the spectra with absorbed single temperature collisionally-ionized diffuse gas model (VAPEC).
We also consider possible broadening of the emission line by multiplying gsmooth function.
The overall expression of emission model is TBabs$\times$ gsmooth$\times$ VAPEC.

The parameters used for fitting are listed in Table \ref{bestfit} for VPSHOCK and VAPEC models.
The elemental abundances for N, O, Ne, Mg, and Fe, whose bright lines are evident in our X-ray spectra, are allowed to vary freely.
Each parameter were initially frozen and then thawed one by one after checking that the previous fit has been improved by thawing, in order to avoid unrestrained-and-unphysical solution.
In the analysis using the VPSHOCK model, the elemental abundances for He, C, S, Ar, Ca, and Ni also need to be adjusted.
However, there are no significant detection of lines from these elements in our X-ray spectrum.
Therefore, we set the abundance of He and C as solar abundance.
Other elements are linked to those of the closest element number, namely, the abundances of S, Ar, and Ca were assumed to be the same as that of Si, and that of Ni was assumed to be the same as that of Fe, since former elements are produced in O burning stage, and latter are related to Si burning stage.
Such treatment does not affect the determination of abundance ratio of N, O, Ne, Mg, Si, and Fe.
Other elements were fixed to the solar values of \citet{Lodders_2009}.  
Details of the treatment are shown in Table \ref{bestfit}.

\begin{figure}
    \centering
    \includegraphics[width=8cm,clip]{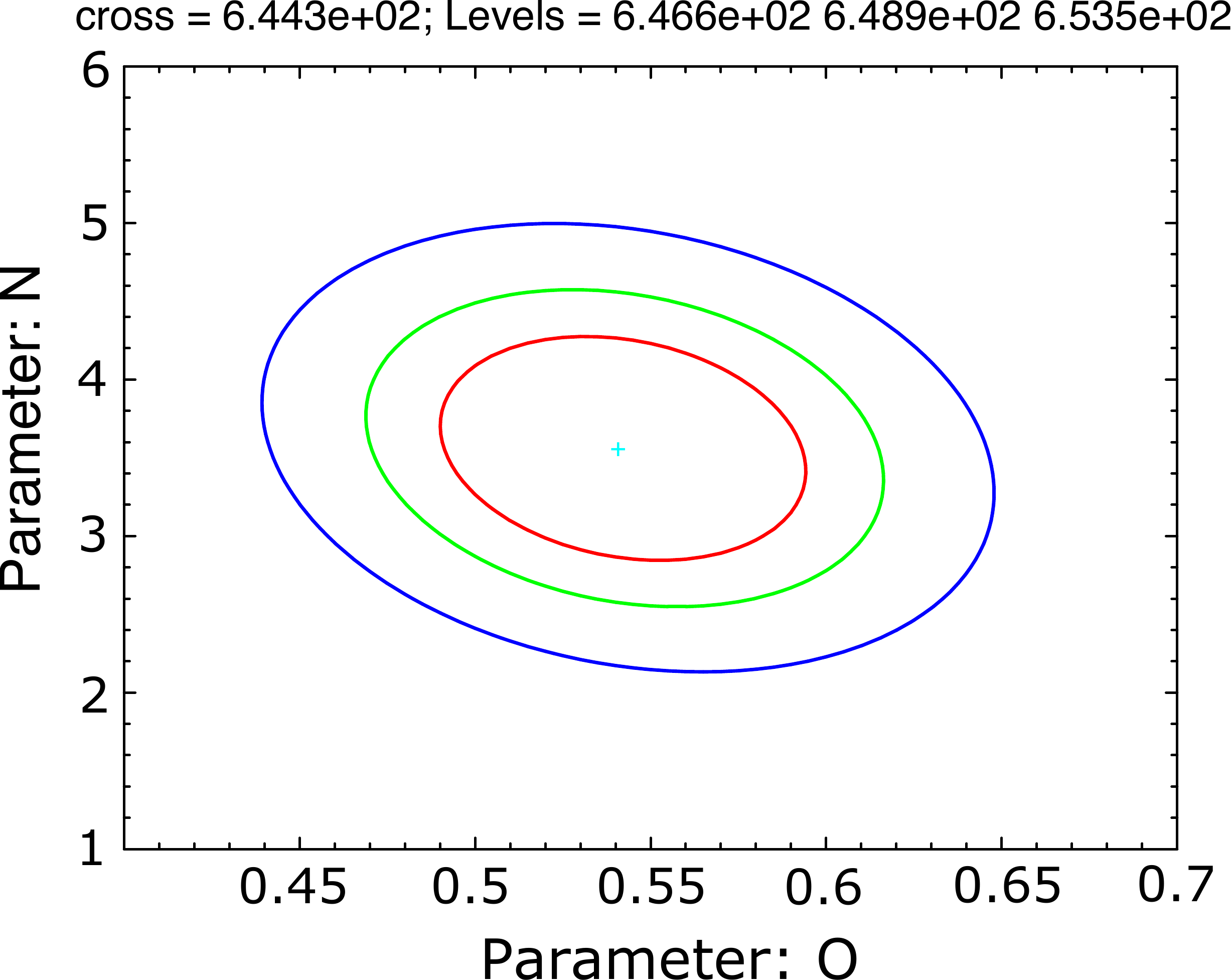}
    \caption{Confidence contour of N and O abundance. 
    The red, green, and blue line represent $68\%, 90\%$ and $99\%$ confidence level, respectively.}
    \label{fig:confidence}
\end{figure}

The best-fit results are shown in Table~\ref{bestfit} and Figure \ref{fig:best_fit}.
Since the VPSHOCK model has smaller reduced chi square value than the VAPEC mode, we used the results obtained from VPSHOCK model.
The hydrogen column density of $N_{\rm H} =4.4 \pm 0.2\times 10^{21}\, \mathrm{atoms\, cm^{-2}}$ and the photon index of $\Gamma =3.0\pm{0.5}$ agree with those obtained at the surrounding regions by past observations \,\citep{Sano_2015}.
The chemical abundances are obtained to be ${\rm N/H} = 3.5\pm0.8{\rm \left(N/H\right)_{\odot}}$, ${\rm O/H} = 0.5\pm0.1{\rm \left(O/H\right)_{\odot}}$, ${\rm Ne/H} = 0.9\pm0.1{\rm \left(Ne/H\right)_{\odot}}$, ${\rm Mg/H} = 
1.0\pm0.1{\rm \left(Mg/H\right)_{\odot}}$, ${\rm Si/H} = 1.0\pm0.2{\rm \left(Si/H\right)_{\odot}}$, and ${\rm Fe/H} = 1.3\pm0.1{\rm \left(Fe/H\right)_{\odot}}$.
Because relative abundances among heavy elements are generally better constrained than absolute abundances $\rm \left(X/H\right)$, we calculate confidence contours between N and O abundances, as shown in Figure \ref{fig:confidence}.  
Based on this result, we estimate the abundance ratio between N and O to be $\rm N/O = 6.8_{-2.1}^{+2.5}\left(\rm N/O\right)_{\odot}$ at $90\%$ confidence level.  
After all parameters are well fitted, we fixed all the parameter except for the redshift, in order to estimate the Doppler velocity of K1.
We also checked that the Doppler velocity obtained solely with the RGS is consistent with that with both the RGS and the MOS as in Table~\ref{bestfit}.

In addition, if we assume that K1 is a sphere of radius of $33''$,  and that the density of electrons and protons is the same, 
from emission measure, we obtain the density and mass of K1 as $4.06_{-0.21}^{+0.27}\, \mathrm{cm^{-3}}$ and $1.72_{-0.09}^{+0.12}\times 10^{-3}\, \mathrm{M_\odot}$, respectively.

\begin{deluxetable}{ccc}
\tablenum{2}
\label{tab:counts}
\tablecaption{
Number of counts used in spectral analysis.
}
\tablewidth{0pt}
\tablehead{\colhead{Instrument} & \colhead{Energy Range} & \colhead{Number of Counts}}
\startdata
\hline
MOS1 & 0.45-8.0\,keV & $4.7\times 10^{3}$\\
MOS2 & 0.45-8.0\,keV & $4.1\times 10^{3}$\\
RGS1 & 0.45-2.0\,keV & $9.1\times 10^{2}$\\
RGS2 & 0.45-2.0\,keV & $1.3\times 10^{3}$\\
\hline
\enddata
\end{deluxetable}

\begin{deluxetable*}{clcc}
\tablenum{3}
\label{bestfit}
\tablecaption{
Spectral-fit Parameters
}
\tablewidth{0pt}
\tablehead{\colhead{Model} & \colhead{Parameter} & \colhead{VPSHOCK} & \colhead{VAPEC}}
\startdata
\hline
TBabs (Local background)           & $N_{\rm H} (\times 10^{22}\ {\rm atoms\ cm^{-2}})$                             & $0.49_{-0.06}^{+0.08}$        & $0.49$ (Fixed)                \\
\hline
power-law                          & Photon index                                                                   & $3.0\pm0.5$                   & $3.0$ (Fixed)           \\
(Local background)                 & ${\rm Norm} \left(\times 10^{-4}\ \mathrm{photons\,keV^{-1}\,cm^{-2}\ @1 keV}\right)$  & $5.2_{-0.7}^{+1.0}$   & $5.4\pm0.8$                      \\
\hline
TBabs (K1)                        & $N_{\rm H} (\times 10^{22}\ {\rm atoms\ cm^{-2}})$                             & $0.53\pm0.02$                 & $0.86_{-0.05}^{+0.04}$                 \\
\hline
gsmooth                            & Gaussian sigma at 1 keV (eV)                                                   & $0.28\pm0.09_{-0.56}^{+0.57}$\tablenotemark{\rm *} & $0.29_{-0.08-0.56}^{+0.09+0.57}$\tablenotemark{\rm *}  \\
(K1)                               & Power of energy for sigma variation                                            &   0 (Fixed)                   & 0 (Fixed)                             \\
\hline
Thermal emission                   & Plasma temperature $k_{\rm B}T_{\rm e}$ (keV)                                  & $0.65\pm0.02$                 & $0.23_{-0.01}^{+0.02}$                \\
 (K1)                              & ${\rm H/H_\odot}$                                                              & 1 (Fixed)                     & 1 (Fixed)                             \\
                                   & ${\rm \left(He\, (=C)/H \right)/ \left(He\, (=C)/H \right)_\odot}$             & 1 (Fixed)                     & 1 (Fixed)                             \\
                                   & ${\rm \left(N/H \right)/ \left( N/H \right)_\odot}$                            & $3.5\pm0.8$                   & $70_{-30}^{+71}$                    \\
                                   & ${\rm \left(O/H \right)/ \left( O/H \right)_\odot}$                            & $0.5\pm0.1$                   & $5.5_{-3.1}^{+20}$                   \\
                                   & ${\rm \left(Ne/H \right)/ \left( Ne/H \right)_\odot}$                          & $0.9\pm0.1$                   & $6.9_{-3.7}^{+21}$                   \\
                                   & ${\rm \left(Mg/H \right)/ \left( Mg/H \right)_\odot}$                          & $1.0\pm0.1$                   & $6.2_{-3.7}^{+21}$                    \\
                                   & ${\rm \left(Al/H \right)/ \left( Al/H \right)_\odot}$                          & -                             & 1 (Fixed)                    \\
                                   & ${\rm \left(Si\,(=S=Ar=Ca)/H \right)/ \left( Si\,(=S=Ar=Ca)/H \right)_\odot}$  & $1.0\pm0.2$                   & $19_{-8.1}^{+62}$                   \\
                                   & ${\rm \left(Fe\, (=Ni)/H \right)/ \left(Fe\, (=Ni)/H \right)_\odot}$           & $1.3\pm0.1$                   & $8.5_{-1.8}^{+43}$                   \\
                                   & Lower limit on ionization timescale $\left(\times 10^{11}\ {\rm s\ cm^{-3}}\right)$ & 0 (Fixed)                & -                             \\
                                   & Upper limit on ionization timescale $\left(\times 10^{11}\ {\rm s\ cm^{-3}}\right)$ & $1.13_{-0.11}^{+0.12}$   & -                \\
                                   & Redshift $\left(\mathrm{km\,s^{-1}}\right)$                                    &  $-230_{-35}^{+49}$           & $-28_{-46}^{+25}$                    \\
                                   & Emission measure $\left(\times 10^{19}\, \mathrm{cm^{-5}}\right)$              &  $1.62_{-0.16}^{+0.23}$       & $3.87_{-3.51}^{+32.7}$                \\
\hline
$\chi^{2}/{\rm d.o.f}$             &                                                                                &$1.26 \left(=663.285/525\right)$& $1.41 \left(=722.10/512\right)$ \\
\hline
\enddata
\tablenotetext{}{
{\bf Note.} The best fit parameters of the spectra fitting of the {\it XMM-Newton}/RGS (0.45-2.0\,keV) and MOS (0.45-8.0\,keV).
The errors represent 90\% confidence level on an interesting single parameter.\\
\tablenotemark{\rm *}: The 1st error represent statistical error, while the 2nd represent systematic error in 1$\sigma$ confidence level\,\citep{Herder_2002}.
}
\end{deluxetable*}

 
\subsection{Time Variability of K1}
It is interesting to note that K1 was already detected in the {\it ROSAT} era and was the second brightest source to be detected after the compact object. 
Using the {\it ROSAT}/PSPC observation ($T_\mathrm{obs}$=2.7 ks) performed in 1992 and the four {\it XMM-Newton} observations ($T_\mathrm{obs}$=11 ks, 17 ks, 14 ks and 43 ks) that covered this region, we investigated a possible time variability of the source over 25 years. 
A common extraction region centered on K1 and of radius $45''$ was used to derive an energy flux in the common 0.6-2 keV energy band. 
The same spectral model (described in Table \ref{bestfit}) was used for all observations and only the normalisation was allowed to vary. 
The background was estimated in each observation in an annulus region surrounding K1 of size $55''<R<200''$. 
The background spectrum is then subtracted in the fitting process.
While the source region K1 is observed near the optical axis in 2007 and 2017, it is at the edge of the field of view in the 2001 and 2004 observations at an off-axis angle of $\sim12'$.
Comparing the flux for these different observations requires a good calibration of the vignetting effect at large off-axis angles. 
The Section 4.5 of the XMM-Newton technical note \textit{CAL-TN-0018}\footnote{\url{https://xmmweb.esac.esa.int/docs/documents/CAL-TN-0018.pdf}} mentions differences in flux for off axis sources of $\sim$5\% which we consider as our systematic uncertainty.

\begin{figure}
    \centering
    \includegraphics[width=9cm,clip]{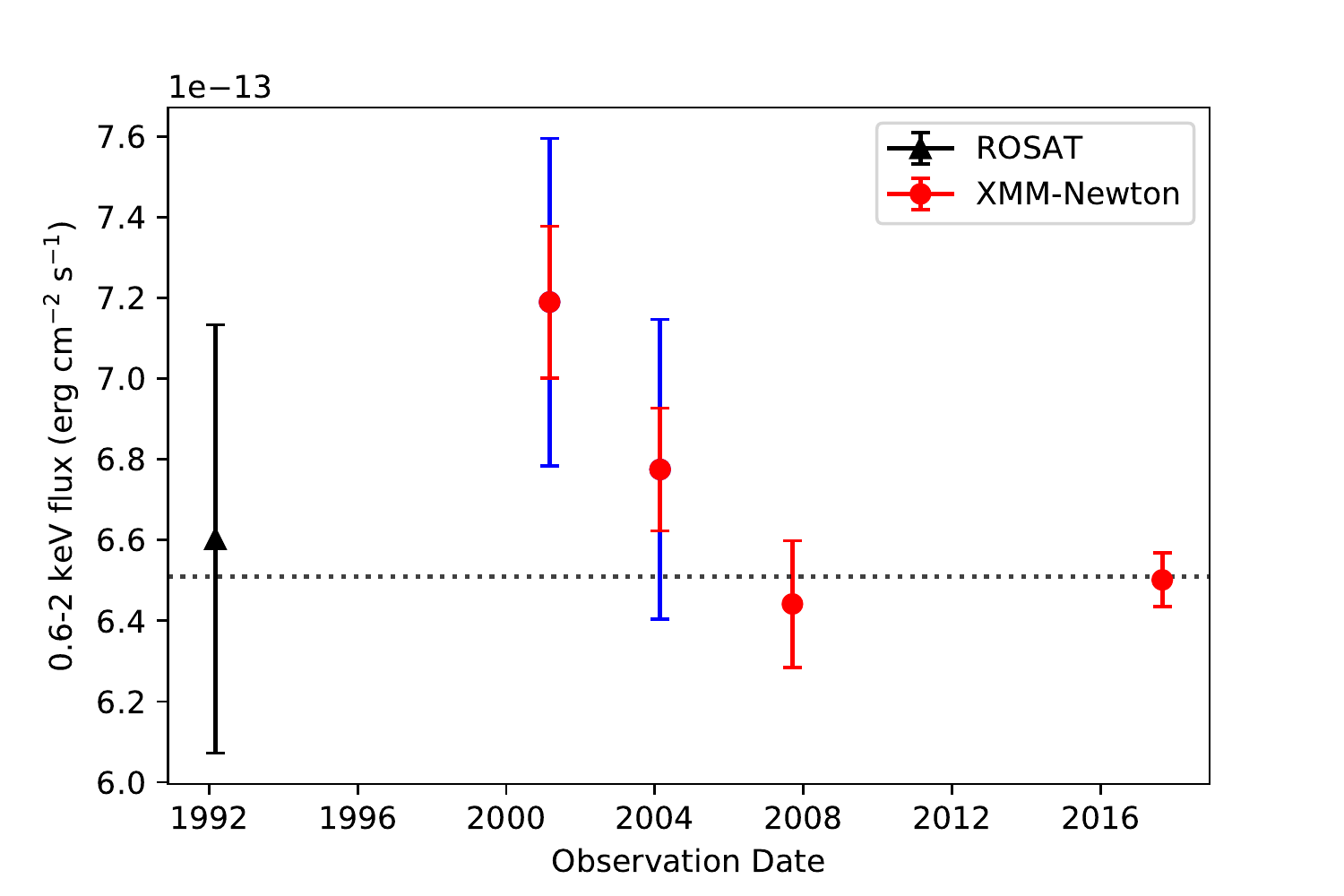}
    \caption{
    Light curve of K1 feature over a 25 year period using a common extraction region and a common spectral model for all observations. 
    The dotted line represents the best-fitted constant flux.
    Statistical error bars are reported at the $1\sigma$ level and a systematic error of 5\% is added for the 2001 and 2004 observations where the source was observed near the edge of the camera. The blue error-bars represent total error for the 2001 and 2004 observations.}
    \label{fig:rosat_compair}
\end{figure}

The resulting light curve is shown in Figure \ref{fig:rosat_compair} with statistical errors only and statistical plus a 5\% systematic error added in quadrature (only for the 2001 and 2004 data).
In order to quantitatively evaluate the time variability of flux of K1, we fitted a constant flux model to the light curve without and with systematic errors and obtained a $\chi^{2}$ of 14.32 and 3.54 respectively for 4 degrees of freedom. 
This corresponds to a rejection of the constant flux model at a $2.7\sigma$ and $0.7\sigma$ level respectively.
We conclude that the flux of K1 does not fluctuate significantly with time and that the slight flux offset of 2001 and 2004 is likely impacted by systematic effects of vignetting calibration.
This variability study therefore helps rules out a background transient source.

\section{Discussion} \label{subsec:tables}

We found an interesting knot emitting thermal X-ray emission in the Galactic SNR RX~J1713.7-3946.  We revealed its detailed X-ray morphology and measured the elemental abundances as well as its Doppler velocity, based on high-resolution X-ray spectrum with the RGS.  Below we discuss the origin of K1 by using these results.  We will also discuss its implication for the initial mass of the progenitor star that produced RX~J1713.7-3946.

\subsection{Origin of K1}

\begin{figure}
    \centering
    \includegraphics[width=8cm,clip]{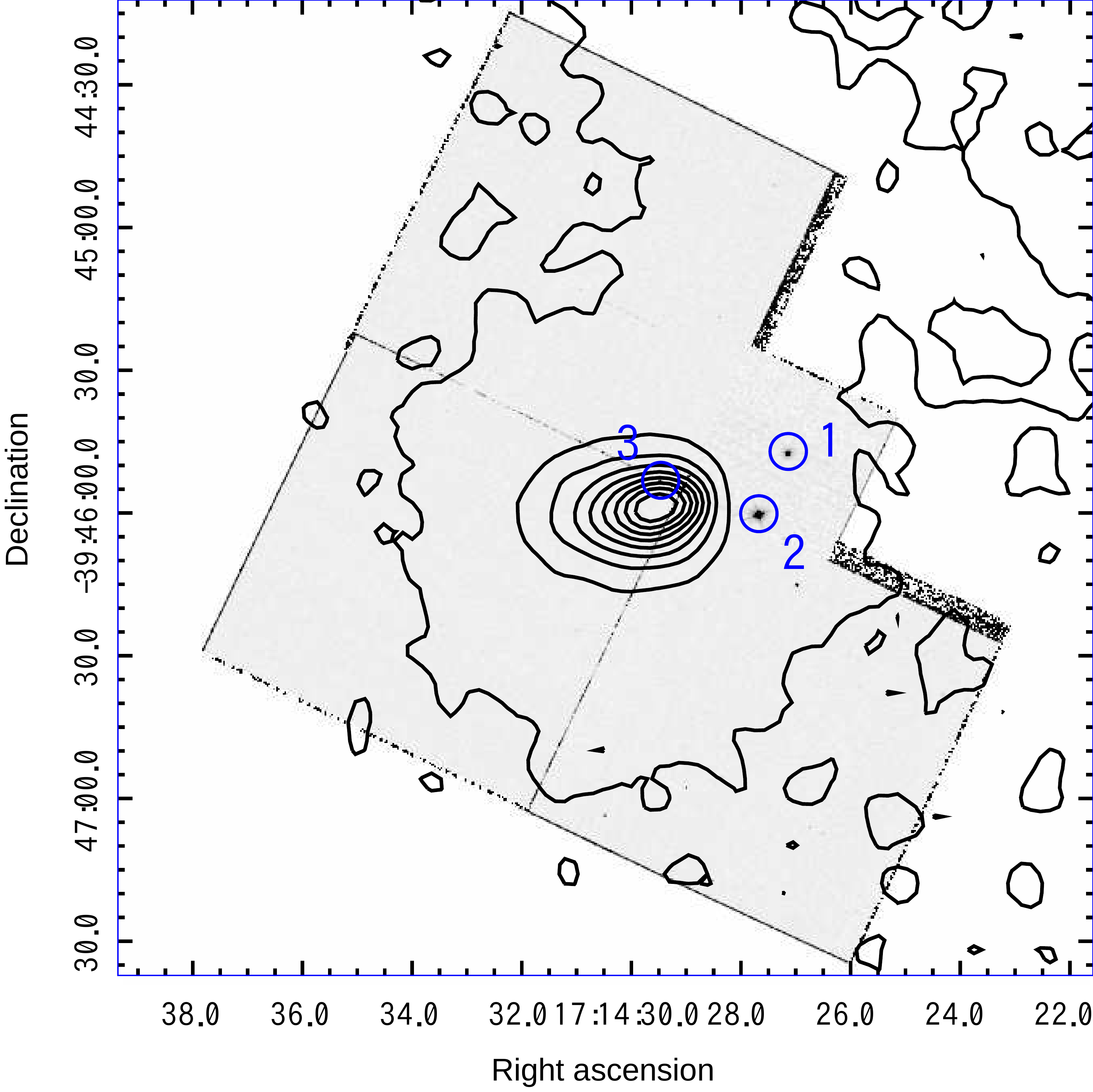}
    \caption{
    Optical image of the area around K1 observed with {\it HST}.
    552 \AA, 433 \AA, and 336 \AA\, data are combined to make this image.
    Contour indicates K1 observed with MOS. 
    The 1, 2, and 3 in the figure represent star names of WR 85, V915 Scorpii, and Gaia\ EDR3\ 5972220096028907008, respectively.
    }
    \label{fig:knot_optical}
\end{figure}

We first searched for optical counterparts, by comparing the X-ray image with an optical image obtained with {\it Hubble Space Telescope}/Wide Field and Planetary Camera 2. 
Figure \ref{fig:knot_optical} shows the {\it HST} image around K1 with X-ray contours overlaid.  
We found 3 stars within or close to K1, as indicated by blue circles in the figure, for which we further investigate relationships with K1:1.\ WR\ 85, 2.\ V915\ Scorpii, 3.\ Gaia\ EDR3\ 5972220096028907008.

As we described in the previous section (see Figure \ref{fig:radial}), K1 extends larger than the PSF of the MOS, which is statistically significant in $5\sigma$. 
From this result, it is clear that K1 is not a single star.  
Rather, its chemical composition (N enhancement) as well as the velocity (Doppler velocity of $\sim$200\,km\,s$^{-1}$) suggests that it is circumstellar material, i.e., debris of stellar winds.  
We will discuss whether these 3 candidate objects create X-ray nebulae like K1 of our interest.

\subsubsection{WR~85}
There are mainly two types of nebulae formed by Wolf-Rayet stars: 1) the pinwheel nebulae, which is caused by the collision of the stellar winds of 
Wolf-Rayet and OB-type stars, and 2) the ring nebulae, which is broadly assumed to be the wind-blown bubble.
The diameter of the pinwheel nebulae was found to be 2.76 mpc and 2.23 mpc 
in WR 98a\,\citep{wr98} and WR 104\,\citep{WR_pin}, respectively.
Given that the distance to WR~85 is $2.1_{-0.2}^{+0.3}$ kpc from the Earth, based on parallax measurements with the Gaia satellite, the physical diameter of K1 is about 0.61 pc, which is inconsistent with the results expected from the pinwheel nebulae.

Next, we consider the shape and position of K1.
If we assume that K1 is the nebula of WR~85, then its shape becomes highly asymmetric.
As already described in Section \ref{sec:obs}, the center of K1 is inconsistent with WR~85.
These results are unfavorable for the wind-blown bubble around Wolf-Rayet nebulae.
In addition, we compared the size of K1 with those of Wolf-Rayet ring nebulae observed in optical and X-ray.
The objects considered are shown in Table \ref{tab:wr_ring}.
From Table \ref{tab:wr_ring}, the optical and X-ray observations suggest that a Wolf-Rayet ring nebula has a diameter of $1\mathchar`- 70$ pc.
Therefore, the ring nebula is more than 1.64 times larger than the diameter of K1 (at a distance of 2.1 kpc), arguing against the possibility that K1 is a ring nebula around WR~85.

\begin{deluxetable*}{lcccc}
\tablenum{4}
\label{tab:wr_ring}
\tablecaption{
Comparison of the size of WR ring nebula by optical and X-ray observations.
}
\tablewidth{0pt}
\tablehead{\colhead{Nebula}&\colhead{Name of WR Star} & \colhead{Mass of WR Star $\left(\rm M_{\odot}\right)$} & \colhead{Diameter in Optical(pc)} & \colhead{Radius in X-ray (pc)}}
\startdata
            \hline
            S 308 & WR 6                 &23\tablenotemark{\rm a}& 17.5   
                         & 8.8\\
            NGC 2359 & WR 7              &13\tablenotemark{\rm a}& 6.54   
                         &  - \\
            NGC 3199 & WR 18               &38\tablenotemark{\rm a}& 15.4-19.2                       & $>7$\\
            MR 26 & WR 22                 &68\tablenotemark{\rm a}& 10.9-25.5                       & 3.2\\
            RCW 58 & WR 4                &28\tablenotemark{\rm a}& 6.11-7.85                       &  - \\
            $\theta$ Mus & WR 48          &-                      & 11.8-20.9                       &  -\\
            MR 46 & WR 52                 &$8.5_{-0.5}^{+0.6}$\tablenotemark{\rm c}& 23.3-34.9      &  -\\
            RCW 78 & WR 55                &14\tablenotemark{\rm a}& 52.4-73.3                       &  -\\
            RCW 104 & WR 75               &18\tablenotemark{\rm a}& 3.49-7.85                       &  -\\
            G 2.4+1.4 & WR 102             &$16.1_{-1.4}^{+1.7}$\tablenotemark{\rm c}& 27.9-31.4     &  -\\
            M1-67 & WR 124                 &22\tablenotemark{\rm a}& 0.91 
                           &  -\\
            MR 95 & WR 128                 &5-11\tablenotemark{\rm b}& 32.0                          &  -\\
            L 69.8+1.74 & WR 131           &39\tablenotemark{\rm a}& 3.67 
                           &  -\\
            MR 100 & WR 134                &18\tablenotemark{\rm a}& 10.4 
                           &  -\\
            NGC 6888 & WR 136              &23\tablenotemark{\rm a}& 4.19-6.28                       & 3.2\\
            \hline
\enddata
\tablenotetext{}{
{\bf Notes.} 
Optical observations were taken from \citet{wr_ring} and the reference there in.
X-ray observations were taken from \citet{wr_bubble}.
The distance used to determine the diameter in visible light observations 
is taken from\,\citep{Hucht_1981}.
\tablenotemark{a},\tablenotemark{b},\tablenotemark{c}, represents data from \citet{van_2001}, and \citet{hamann_2019}, \citet{sander_2019} respectively.
}
\end{deluxetable*}

\begin{figure}[htbp]
    \begin{center}
    \includegraphics[width=8cm,clip]{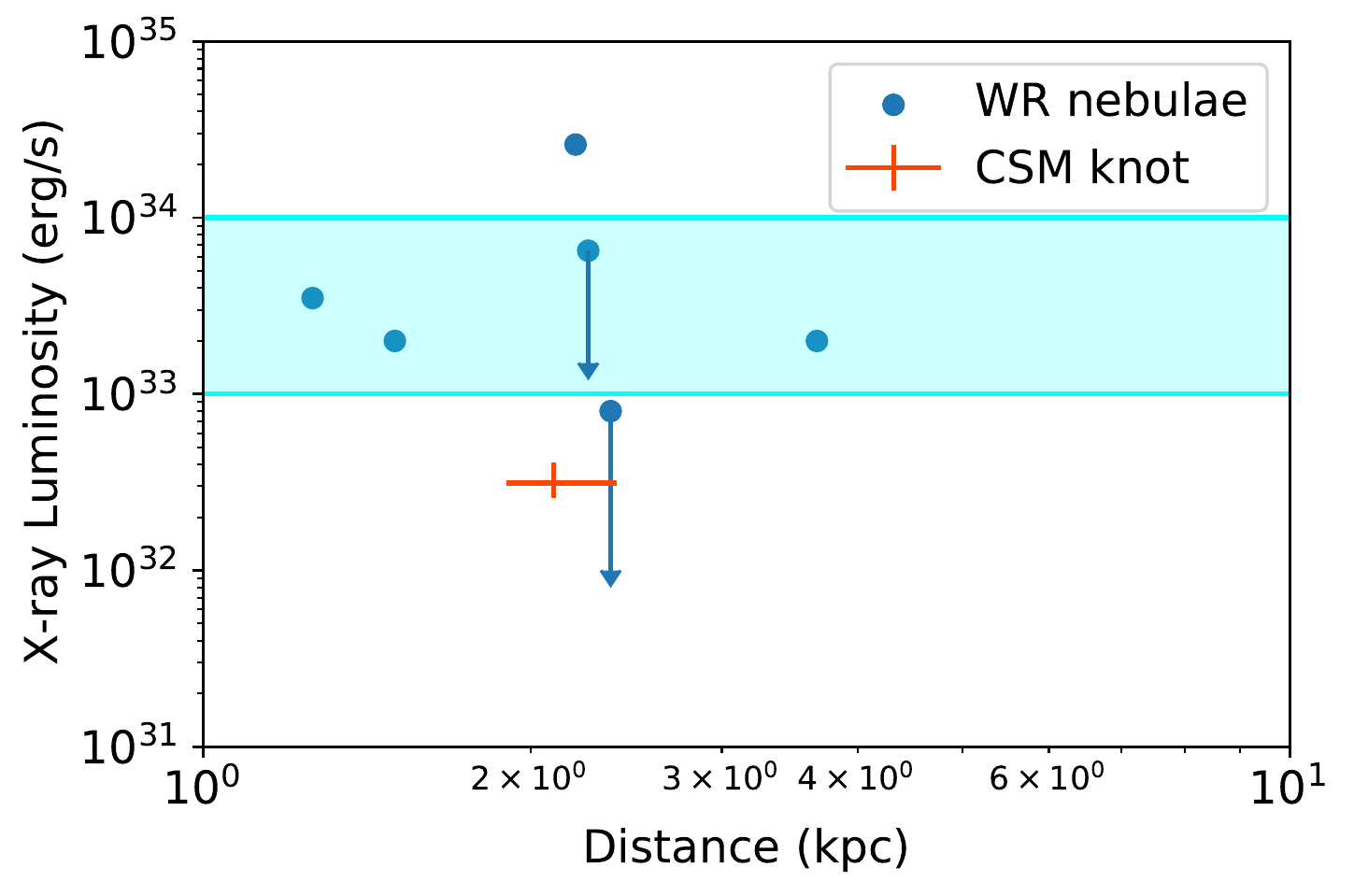}
    \caption{
        Comparison of X-ray luminosity between K1 and Wolf-Rayet nebulae.
        The blue dots indicate X-ray luminosity \,\citep{wr_bubble,wr_bubble2,Toala_2017} of Wolf-Rayet nebulae,
        and the orange cross line indicates the X-ray luminosity from K1 at the WR~85 assumed distance of $2.1_{-0.2}^{+0.3}$\,kpc.
        WR~7 shows the luminosity at 0.3-2.0\ keV, WR~8 at 0.3-3.0\ keV, and the rest at 0.3-1.5\ keV.
    }
    \label{fig:wr_vs_knot}
    \end{center}
\end{figure}

We also compared the X-ray ($0.3 \mathchar`- 1.5 {\rm keV}$) luminosities 
of K1 and Wolf-Rayet nebulae.
Figure \ref{fig:wr_vs_knot} shows X-ray luminosity of Wolf-Rayet nebulae as a function of distance from the Earth\,\citep{wr_bubble,wr_bubble2}, together with that of K1 as a cross.
The X-ray luminosity of WR nebulae is in the range of $10^{33} \mathchar`- 10^{34}\, \mathrm{erg\,s^{-1}}$\,\citep{Chu_2003,Toala_2012,wr_bubble2,Toala_2016}.  
If we assume that K1 is located at $2.1_{-0.2}^{+0.3}\ {\rm kpc}$, which is consistent with WR~85, the observed X-ray luminosity becomes $\lesssim 0.31$ times larger than the typical luminosity of WR nebulae, which is in tension with the relation between K1 and the WR~85.  
Given these considerations, we conclude that K1 is not a nebula created by the mass loss of WR~85.

\subsubsection{V915 Scorpii}
V915 Scorpii is a Yellow Supergiant (YSG)\,\citep{scopi} or a Yellow Hypergiant (YHG) \,\citep{hyper} with a G5 Ia type spectrum.
It is located $21.60''$ away from K1, and its distance is 
estimated to be $1.8\pm0.2$ kpc from the Earth with Gaia.

\citet{Smith_2014} estimated the stellar wind velocities due to mass loss 
at each stage of stellar evolution to be 20--40 km\,$\mathrm s^{-1}$ and 30--100 km\,$\mathrm s^{-1}$ for YSG and YHG, respectively.
Assuming a stellar wind velocity of 100\,km\,$\mathrm s^{-1}$ from the star, temperature of the circumstellar gas heated by the stellar wind would 
be $1.07\times 10^{-6}$\ keV according to the Rankine-Hugoniot equation.
This is too cool to radiate significant X-rays.
Therefore, we conclude that slow stellar wind of V915 Scorpii is unfeasible origin of K1. 

\subsubsection{Gaia~EDR3~5972220096028907008}
This star is located $4.89''$ away from K1.
Its distance is estimated to be $2.4\pm{0.1}$\ kpc from Gaia observations.
Observations in the G, ${\rm G_{BP}}$, and ${\rm G_{RP}}$ bands by the Gaia satellite show that the star has ${\rm G_{BP}} - {\rm G_{RP}} = 1.15$\ mag and the absolute magnitude in the G band is ${\rm M_{G}} = 1.83$ 
mag.
By comparing this result with the HR diagram of the star observed by the Gaia satellite\,\citep{Gaia_HR}, it was suggested that this star is a subgiant star with a surface temperature of about 5,000 K.
Since it is difficult for subgiant stars to produce X-ray emitting nebulae, we concluded that this star is not the origin of K1, either.  

\subsubsection{Possibility of K1 originate from stellar object which has no relation to SNR~RX~J1713.7-3946}
In summary, there is no promising stellar candidate that can create K1.
In the following we also discuss the possibility of K1 originate from cluster of galaxies, pulsar wind nebulae, binary star nebulae, and SNR shell.

We can use the redshift to distinguish between clusters of galaxies and galactic objects.
The closest cluster of galaxies to earth is Virgo cluster, located about 16.5\, Mpc away from Earth\,\citep{Simona_2007}.
The redshift of virgo cluster is 1132 km/s, which is higher than our measurement by a factor of ~5.
Therefore, we concluded that cluster of galaxies is not the origin of K1.

The wind from a pulsar will create a nebula around itself, called pulsar wind nebula\, \citep{Kargaltsev_2015}.
The fast wind from the pulsar will collide with CSM, creating shock, which accelerate electrons and protons.
For this reason, the X-ray from pulsar wind nebulae will be dominated by non-thermal emission, which is not suitable to our analysis result.
Therefore, we concluded that a pulsar wind nebula is not the origin of K1.

The binary star will create planetary nebulae around itself\, \citep{Jones_2017, Boffin_2019}.
From X-ray observations with {\it Chandra}, it is known that some of them emit point and/or diffuse X-rays\,\citep{Hoogerwerf_2007, Kastner_2012}.
However, no binary stars are found inside K1.
Just outside K1, there is a very bright star, V915 Scorpii, which is also known to be a single star.
Therefore we conclude that binary star nebulae are not the origin of K1.

Lastly, we discuss the possibility of K1 originate from an SNR which does not have relation with RX~J1713.7-3946.
By analysing SNRs in the LMC, \citet{Ou_2018} reported that X-ray (0.3-8.0\, keV) luminosity will be in the range of $10^{34}$ to $10^{38}$ erg/s.
Our measured X-ray flux of K1, combined with the luminosities of LMC SNRs, places it to a distance of $1.1\times 10^{-2}$ - $1.1$~Mpc.
This means that K1 is located outside our Galaxy, and conflicts with the fact that the hydrogen column density for K1 is small enough to be within our Galaxy.
From this reason, we conclude that the SNR is not the origin of K1.

\subsubsection{Possible Origin of K1}
With the discussion above, there is no promising stellar object which can create K1.
On the other hand, the hydrogen column densities around K1, which is measured by past X-ray observations\,\citep{Sano_2015} are comparable to that obtained at K1 $\left(N_{\rm H} = \left(0.53 \pm 0.02\right) \times 10^{22}\ {\rm atoms\, cm^{-2}}\right)$.
This result suggests that both K1 and the SNR are located at the same distance and K1 should be associated with the SNR.
In addition, as it shows in Figure \ref{fig:ellipse}, the major axis of ellipse points toward the center of the SN.
This result suggest that K1 was heated by the shock of the SN, which also support the association of the SNR.
Therefore, with the nitrogen enhanced chemical abundance, we argue that K1 is most likely the shock-heated CSM of the progenitor star that produced RX~J1713.7-3946.

One big mystery in this study is the reason why only one knot in the SNR emit thermal X-rays.
To answer this problem, we estimated when K1 was ejected from the progenitor star.
By dividing ionization timescale by electron density, we acquired the time scale since K1 was heated to be $\sim 880(nt/1.13\times 10^{11}\,{\rm s\,cm^{-3}})(n_{\rm e}/4.06\,{\rm cm^{-3}})^{-1}$ yr.
The timescale suggest that it was heated recently (about 880 years ago).
However, this timescale is highly dependent on volume of K1, which is uncertain.
One of the possibilities is that the volume of K1 is larger than our assumption, which will lower the density and increase the timescale.
In this case, K1 was located in the vicinity of the progenitor star, suggesting that it was ejected just before the explosion.
Observations of Type~\mbox{I\hspace{-.1em}IP} SN suggest that the mass-loss rate increases just before the explosion\,\citep{Moriya_2017}, and which may create a dense CSM.
It has also been suggested that the CSM ejected just before the explosion is asymmetrically distributed\,\citep[e.g.,][]{Andrews_2017}.
Episodic asymmetrical mass ejection just before the SN explosion may explain why only one knot structure exist in the SNR.
Further observations are required to elucidate the details, though.

\subsection{Ion Temperature derived from Line Widths}\label{sec:temperature}
Assuming that the widths of emission lines (as in Table \ref{bestfit}) are broadened only by the thermal motion of ions, we can infer the ion temperature.
It is reasonable to consider that the line width is determined mainly by the O~Ly$\alpha$ line, as it is the strongest and cleanest line in the RGS spectrum.
The width of O~Ly$\alpha$ line is $0.18\pm0.06_{-0.23}^{+0.26}$~eV.
The 1st and 2nd error represent statistic and systematic error, respectively.
For the systematic error, we adopted $\pm7\, {\rm m\AA}$, which is random systematic wavelength error of RGS in $1\sigma$\,\citep{Herder_2002}.
From this width, we derive $k_{\rm B}T_{\rm O}$ to be $1.15_{-0.11-0.94}^{+0.13+5.33}$\, keV.
This temperature is broadly consistent with the electron temperature of $k_{\rm B}T_{\rm e} = 0.65\pm 0.02$\, keV, suggesting temperature equilibrium between ions and electrons.

Since we obtain that electron density $\left(n_\mathrm{knot} = n_{\rm p, e} = 4.06\,\mathrm{cm^{-3}}\right)$ is about 400 times larger than that of interstellar medium near the SNR\,$\left(n_\mathrm{ISM} = 0.01\, \mathrm{cm^{-3}}\right)$, it is reasonable to assume that K1 is heated by slower shock which propagate into CSM (i.e. transmitted shock) immediately after the SN, whose velocity is lower than the blast-wave propagating into its surrounding medium.
If we use the density and velocity in K1 and ejecta as $n_\mathrm{knot} = 4.06$\, $\mathrm{cm^{-3}}, n_\mathrm{ISM} = 0.01\, \mathrm{cm^{-3}}, v_\mathrm{transmitted}$, and $v_\mathrm{blastwave}\approx3000\, {\rm km\, s^{-1}}$\citep{Tsuji_2016, Fabio_2017, Tanaka_2020}, respectively, from pressure equilibrium $\left(n_\mathrm{knot}v_\mathrm{transmitted}^{2} = n_\mathrm{ISM}v_\mathrm{blastwave}^{2} \right)$ we obtain $v_\mathrm{transmitted} \approx 150\, {\rm km\, s^{-1}}$.
If K1 was heated by this shock, then the ion temperature for O will be 0.70 keV, which is not significantly different from the observed electron temperature of K1.
This temperature is also statistically compatible with O temperature derived from line width.

\subsection{Mass Estimation of the Progenitor Star of RX~J1713.7-3946}
We use the chemical composition of the possible CSM knot to infer the mass of the progenitor star.
To this end, we compared the N/O ratio obtained from our X-ray spectral analysis with those of the outer layers in the red supergiant star expected from stellar evolution simulations.
\begin{figure}[htbp]
    \begin{center}
    \includegraphics[width=8cm,clip]{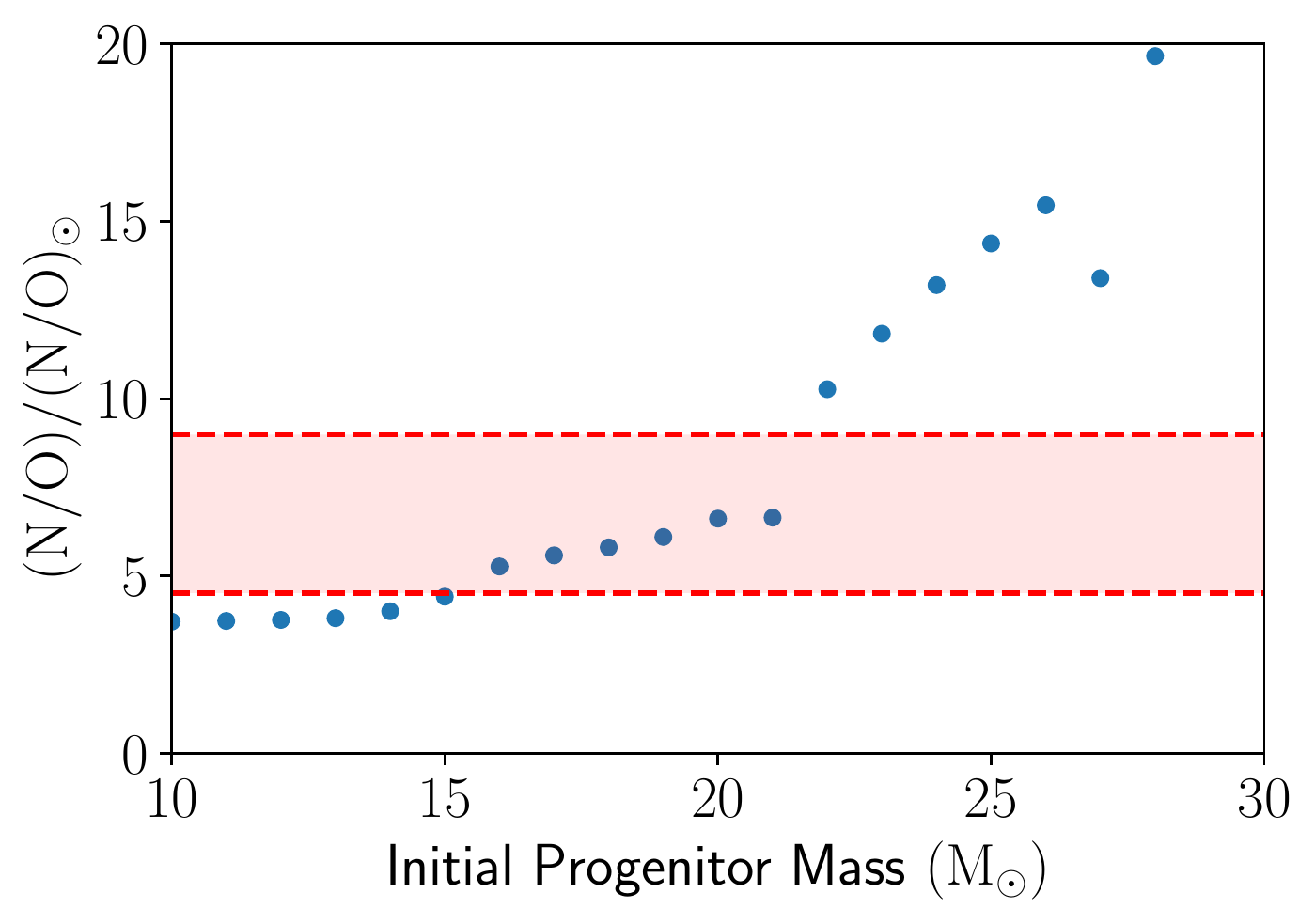}
    \caption{
        Progenitor star mass and $\rm N/O$ ratio in the outer layers of red supergiants.
        The blue dots shows the simulation results (Model L in\,\citealt{Yoshida_2019}), and the red hatched area shows the results obtained by this study.
    }
    \label{fig:no_compair}
    \end{center}
\end{figure}
Figure \ref{fig:no_compair} shows the $\rm N/O$ ratios of the outer layers of the red supergiants as a function of mass of the progenitor star.
The blue dots represent the simulation results (Model L in\,\citealt{Yoshida_2019}), and the red filled area shows the range of our observations.
From this plot, the initial mass of the progenitor star is estimated to 
be $15\, \rm M_{\odot} \lesssim \rm M \lesssim 20\, \rm M_{\odot}$.  
This result agrees with the estimate $\left(\lesssim20\,\rm M_{\odot}\right)$ from the chemical abundance of then ejecta\,\citep{Katsuda_2015}, and the estimate based on a size of the cavity created by the wind from the progenitor star $\left(12\mathchar`- 16\,\rm M_{\odot}\right)$\,\citep{Cassam_2004}.

\section{Conclusion}

We discovered a possible N-rich CSM knot inside the synchrotron-dominated SNR RX~J1713.7-3946.  
The X-ray spectrum obtained with {\it XMM-Newton}/RGS showed clear X-ray line emission, including N Ly$\alpha$, O Ly$\alpha$, Fe XVIII, Ne X, Mg XI, and Si XIII for the first time from this knot.  
The spectral analysis revealed that the abundance of N is $\sim$3.5 times higher than the solar value and those of other elements are near solar values, from which we inferred that it is the CSM ejected when the progenitor star evolved into a red supergiant phase.  
By comparing the abundance ratio of $\rm N/O = 6.8_{-2.1}^{+2.5}\left(N/O\right)_{\odot}$ with those expected in H-rich layers of red supergiant stars, we estimate the initial mass of the progenitor star to be $15\, \rm M_{\odot} \lesssim \rm M \lesssim 20\, \rm M_{\odot}$.  This result agrees with other estimates, i.e., the chemical abundance of the ejecta and the size of the pre-explosion cavity.
The fact that only one thermal knot is observed in the SNR is still a mystery.
We propose that this particular knot might be related to an episodic asymmetrical mass ejection just before the SN explosion, but further observations are required to elucidate this issue.

\acknowledgments
We would like to thank Drs.\ Yasuharu Sugawara, Makoto S.\ Tashiro, Kosuke Sato, Mr.\ Yuji Sunada, and Nobuaki Sasaki for constructive comments.
This work was supported by the Japan Society for the Promotion of Science KAKENHI grant numbers 20H00174 (SK), 21H01121 (SK, YT, SF), 20K03957 (SF), 20K0409 (YT), 20H00158, 20H05249 (TY).  
This work was partly supported by Leading Initiative for Excellent Young Researchers, MEXT, Japan.

\bibliography{sample63}{}
\bibliographystyle{aasjournal}



\end{document}